# Preferential out-of-plane conduction and quasi-one-dimensional electronic states in layered 1T-TaS$_2$

E. Martino*[1,2], A. Pisoni[1], L. Ćirić[1], A. Arakcheeva[1], H. Berger[1], A. Akrap[2], C. Putzke[3,4], P. J.W. Moll[3,4], I. Batistić[5], E. Tutiš[6], L. Forró[1], K. Semeniuk*[1]

[1]*École Polytechnique Fédérale de Lausanne (EPFL), Institute of Physics, CH-1015 Lausanne, Switzerland*

[2]*University of Fribourg, Department of Physics, CH-1700 Fribourg, Switzerland*

[3]*École Polytechnique Fédérale de Lausanne (EPFL), Institute of Materials Science and Engineering, CH-1015 Lausanne, Switzerland.*

[4]*Max Planck Institute for Chemical Physics of Solids, 01187 Dresden, Germany.*

[5]*Department of Physics, Faculty of Science, University of Zagreb, HR-10000 Zagreb, Croatia*

[6]*Institute of Physics, HR-10000 Zagreb, Croatia.*

*email: edoardo.martino@epfl.ch, konstantin.semeniuk@epfl.ch

## Abstract

Layered transition metal dichalcogenides (TMDs) are commonly classified as quasi-two-dimensional materials, meaning that their electronic structure closely resembles that of an individual layer, which results in resistivity anisotropies reaching thousands. Here, we show that this rule does not hold for 1T-TaS$_2$—a compound with the richest phase diagram among TMDs. While the onset of charge density wave order makes the in-plane conduction non-metallic, we reveal that the out-of-plane charge transport is metallic and the resistivity anisotropy is close to one. We support our findings with *ab-initio* calculations predicting a pronounced quasi-one-dimensional character of the electronic structure. Consequently, we interpret the highly debated metal-insulator transition in 1T-TaS$_2$ as a quasi-one-dimensional instability, contrary to the long-standing Mott localisation picture. In a broader context, these findings are relevant for the newly born field of van der Waals heterostructures, where tuning interlayer interactions (e.g. by twist, strain, intercalation, etc.) leads to new emergent phenomena.

## Introduction

Despite remaining at the forefront of research for more than a decade[1], quasi-two-dimensional materials still provide a vast playground for discovery of novel electronic behaviour, showing promise for a variety of technological applications[2,3]. Prominent among the recent advances in the field is the demonstration of how new phenomena can be induced in known compounds by tuning the interlayer interaction, as exemplified by the unconventional superconductivity and strong electronic correlations in magic-angle graphene superlattices[4] or long-lived excitons in heterobilayers of transition metal dichalcogenides (TMDs)[5,6]. Characterisation of the coupling between the layers in van der Waals materials is therefore of great interest, and valuable information can be revealed by probing the out-of-plane charge transport. However, contrary to its in-plane counterpart this property has not been given a lot of attention so far, with the existing data being rather scarce and in some cases inconsistent[7–11]. With the aforementioned discoveries in mind, one can raise a question: to what extent can various layered materials be really classified as quasi-two-dimensional?

1T-TaS$_2$ stands out from the other conducting TMDs with its remarkably rich pressure-temperature phase diagram (Fig. 1a), containing a record number of distinct charge density wave (CDW) orders of diverse electronic properties[12]. The commensurate (C) CDW phase is characterised by the $\sqrt{13} \times \sqrt{13}$ reconstruction (Fig. 1b), which is manifests in a superlattice of David-star-shaped clusters (abbreviated as DS) of 13 Ta and 26 S atoms[13]. At room temperature 1T-TaS$_2$ assumes the so-called nearly-commensurate (NC) CDW order, occurring uniquely in this compound. While the periods of the charge density modulation and the lattice in the NC phase are incommensurate, the deviation from the commensurability is so small that the $\sqrt{13} \times \sqrt{13}$ reconstruction can still take place over a few-nanometer range. As a result, the DS become organised into a nano-array of hexagonal domains, which are separated by regions of aperiodically distorted lattice (discommensurations). In the direction perpendicular to the layers the domains obey the face-centered cubic packing, with a three-layer periodicity[14] (Fig. 1e). Lowering temperature causes the NC to C phase transition, accompanied by a sizeable jump in resistivity. This change can, however, be reversed via an application of light or current pulse, which makes the material a promising candidate for applications in devices[11,15–17]. The pronounced sensitivity of the NC to C phase transition to the sample thickness[16,18] implies that the interlayer interactions play a fundamental role in the process, which further motivates the investigation of the out-of-plane charge transport.

Given the intriguing functional properties of 1T-TaS$_2$[11,15–17], there still exist open questions concerning the electronic structure of the compound. First, there is an active ongoing debate regarding the insulating nature of the C phase, which has traditionally been regarded as a Mott insulator[19]. While such attribution is in line with a number of spectroscopic investigations[20–22], the correlated insulator picture has been challenged by recent band structure calculations[23,24], which relate the formation of the band gap to the interlayer stacking order of the DS[25,26]. Second, the NC phase displays a rather atypical negative temperature coefficient of the in-plane resistivity[12,16,18,27], a solid interpretation of which has not been proposed so far.

We conducted a highly accurate investigation of the in-plane and out-of-plane electrical resistivities ($\rho_\parallel$ and $\rho_\perp$, respectively) of bulk monocrystalline 1T-TaS$_2$ across all the different CDW phases. We reveal that resistivity anisotropy ($\rho_\perp/\rho_\parallel$) in the compound is, astoundingly, of the order of one. Furthermore, we find that the NC phase is a conventional metal in the out-of-plane direction, in stark contrast to the in-plane behaviour, characterised by the negative temperature coefficient of resistivity, which we relate the composite nature of the lattice. Aided by band structure calculations we attribute the interlayer metallicity to a formation of quasi-one-dimensional electronic states extending along the *c*-axis of the crystal. Consequently, we regard the metal-insulator transition as a quasi-one-dimensional instability – a notion favouring the band-insulator nature of the C phase. Our results paint a peculiar picture: due to the unique structure of 1T-TaS$_2$ in the NC phase, its in-plane charge transport is substantially suppressed compared to the more common states of metallic TMDs[7,8], yet simultaneously, the out-of-plane conduction preserves the coherent metallic character, making the compound an effectively three-dimensional conductor with highly anisotropic electronic properties.

## Results

**Anomalously low resistivity anisotropy**

A major hindrance to a systematic study of the interlayer resistivity in quasi-two-dimensional materials is the characteristic morphology of the available single crystals, which typically have very limited extent along the *c*-axis compared to their lateral dimensions[28]. Such a shape makes it difficult to constrain the current flow in a direction strictly perpendicular to the layers. Furthermore, the intrinsic physics can become obscured by effects caused by crystallographic defects or delamination, which are often prevalent due to weak mechanical interlayer coupling. Many of these problems can be overcome by employing focused ion beam (FIB) microstructuring, which has been utilized with great success for probing charge transport in solid-state systems[29] and was essential in making an accurate measurement of $\rho_\perp$ of 1T-TaS$_2$ possible. To prepare a sample (shown in Fig. 2a), a monocrystalline lamella, with its plane parallel to the *c*-axis, was extracted from a larger crystal and shaped into a microstructure which ensured a well-defined measurement geometry, absence of current jetting effects[30] and minimised the mechanical stresses. Measurements of $\rho_\perp$ and $\rho_\parallel$ were conducted simultaneously, using the same microstructure. We developed an optimal sample preparation procedure (described in the Methods section), which minimised the destructive shear stresses experienced by a microstructure upon changing temperature or applying pressure (Supplementary Note 1 and Supplementary Figure 1 describe the damage shear stresses can cause and how the effect can be mitigated).

The previously published values of resistivity anisotropy of 1T-TaS$_2$ in the C and NC phases are in the range of hundreds[10] and thousands[11], which is suspiciously high given the results of recent band structure calculations and angle resolved photoemission spectroscopy experiments indicating a sizeable band dispersion along the $k_z$-axis of the Brillouin zone[31–35]. A plot of our data for $\rho_\perp$ and $\rho_\parallel$ of 1T-TaS$_2$ between 2 K and 400 K (Fig. 2b) presents comparable resistivity values for the two current directions. In the temperature range between 1.8 K and 400 K, the anisotropy varies between 0.6 and 4.3 (Fig. 2c), which is on average more than two orders of magnitude smaller than the published values[10,11]. Transitions from the incommensurate CDW to NC and then to the C phase upon cooling are accompanied by discontinuous reductions of anisotropy. A clear signature of the triclinic phase[36], which is typically much less pronounced in the existing literature, can be seen in the warm-up curve between 230 K and 270 K. However, the most remarkable behaviour is associated with the NC phase, in which $\rho_\parallel$ increases upon cooling but $\rho_\perp$ is clearly metallic.

With the help of finite element simulations (presented in the Supplementary Note 2 as well as Supplementary Figures 2, 3 ,4) we verified that the chosen geometry of the sample is expected to give reliable resistance readings for a range of hypothetical anisotropy values between approximately 0.1 and 100. Furthermore, we analysed the probing geometries that had been used in earlier measurements of the *c*-axis resistivity of 1T-TaS$_2$[10,11] and identified the shortcomings explaining why the previously published data are so different from ours (see Supplementary Note 3 and Supplementary Figures 5, 6, 7, 8).

Given the interlayer stacking of the DS domains in the NC phase (Fig. 1e), we note that any straight path parallel to the *c*-axis passes through a DS domain at least once every three layers. This implies that the out-of-plane metallicity is the intrinsic property of the domains, unless the conduction happens percolatively, via the discommensurations. Further insight can be gained by comparing 1T-TaS$_2$ with two other TMDs, 2H-TaSe$_2$ and 2H-NbSe$_2$, known to possess *c*-axis metallicity from earlier far-infrared optical reflectivity measurements and having homogeneous in-plane structure[7,8]. FIB-prepared samples were used to obtain values of room temperature resistivity for the two principal directions. As can be seen in Fig. 2d, all three materials show comparable $\rho_\perp$ close to 1 mΩcm, which rules out the percolation scenario for the *c*-axis conduction. At the same time, $\rho_\parallel$ of 1T-TaS$_2$ at room temperature is approximately 6 times higher than for 2H-TaSe$_2$ and 2H-NbSe$_2$. This suggests that the anomalously low resistivity anisotropy in the NC phase of 1T-TaS$_2$ is largely a consequence of the enhanced $\rho_\parallel$, whereas the changes in $\rho_\perp$ are secondary in that regard. This situation remotely resembles the case of the quasi-one dimensional conductor BaVS$_3$, where the on-chain correlations strongly enhance intrachain resistivity, but hardly influence the interchain hopping, which manifests as a very low resistivity anisotropy[37].

**Counterintuitive direction dependence of charge transport character**

The NC phase of 1T-TaS$_2$ can be stabilised over a broader temperature with an application of moderate hydrostatic pressure[12,27]. We employed this approach in order to investigate the low temperature charge transport in the NC phase and its reaction to the reduction of interlayer spacing. The measured $\rho_\parallel$ (Fig. 3a) is consistent with the published results[12,27]. Applying pressure of 0.9 GPa resulted in a full suppression of the C phase, revealing that $\rho_\perp$ in the NC phase remains metallic down to the lowest temperatures (Fig. 3b). Further increase of pressure caused the shape of the low temperature part (0 to 30 K) of $\rho_\perp$ go from approximately linear to quadratic (Fig. 3c inset), typical for Fermi-liquid metals, likely as a result of stiffening of the phonon modes and establishment of electron-electron interactions as the dominant source of scattering. Presence of stacking disorder could explain the rather low residual resistivity ratio. In Fig. 3c we plot $\rho_\perp(T)$ and $\rho_\parallel(T)$ for the highest achieved pressure of 2.1 GPa to emphasise the abnormal contrast between the two conduction characters. On cooldown the anisotropy goes from the room temperature value of 2 down to as little as 0.2, contradicting the intuitive quasi-two-dimensional view of the material. In the subsequent sections of this paper we attempt to rationalise these perplexing observations.

**Suppressed in-plane conduction in the nearly-commensurate CDW phase**

Given that the DS domains and the discommensurations in the NC phase are individually intrinsic in-plane conductors, one can naïvely attempt to explain the negative $d\rho_\parallel/dT$ with a parallel resistor model, assuming that the overall conductivity is a sum of two independent contributions. Upon cooling, the geometry of the domain lattice continuously changes from honeycomb-like (side-sharing arrangement) to Kagome-like (corner-sharing arrangement)[38]. Simultaneously, the domains grow and their centres move away from each other, but the discommensuration network always remains continuous (we note that the perfect Kagome-type or honeycomb-type domain lattice pictures, which are often used in the literature[12,15,39], are oversimplifications in the current context). Our powder X-ray diffraction data shows that pressure increase leads to the same transformation as during a warm-up, and we visualise it in Fig. 4 (additional crystallographic data are presented in the Supplementary Note 4 and the Supplementary Figures 9, 10, 11). On lowering temperature, the rotation towards the Kagome-like arrangement constricts the discommensuration channels, reducing their contribution towards the overall conductivity, which might explain its decrease. To verify such model we approximate the aperiodic discommensurations (approximately 30 % of the in-plane area) with the incommensurate phase of 1T-TaS$_2$ ($\rho_\parallel \approx 0.25$ mΩcm at 360 K), and the DS domains with the commensurate phase of 1T-TaSe$_2$ (metallic and isostructural to 1T-TaS$_2$, $\rho_\parallel \approx 1.5$ mΩcm at 300 K)[40]. At room temperature the NC phase has $\rho_\parallel \approx 0.7$ mΩcm, which is consistent with the expectations, but the value of $\rho_\parallel \approx 1.7$ mΩcm at 130 K cannot be explained by the model, since both 1T-TaSe$_2$ and the incommensurate CDW phase of 1T-TaS$_2$ both become substantially more conductive upon cooling[12,40]. We therefore conclude that either the DS domains are intrinsically much more resistive along the in-plane directions than commensurately reconstructed 1T-TaSe$_2$, or an additional scattering mechanism must be behind the negative $d\rho_\parallel/dT$.

We refer to the far-infrared reflectivity data[41], noting that the real part of the in-plane conductivity lacks the low-frequency Drude peak, characteristic for conventional metals[42]. Such suppression of charge transport in the direct current limit happens to be rather common for nanoscale-structured systems[43], where it is attributed to directionally biased boundary scattering, captured by the Drude-Smith model[44]. Alternatively, the aperiodicity of the NC phase prompts a comparison with certain species of quasicrystals, which also exhibit suppression of the Drude peak[45], and where absence of Bloch waves causes $\rho(T)$ to have a negative slope and follow effectively the same shape in quasiperiodic directions, given a typical metallic behaviour along the periodic one[46]. A more precise determination of the microscopic origin of the negative $d\rho_\parallel/dT$ in the NC phase would require a separate study.

**Quasi-one-dimensional states as a source of metallic out-of-plane conduction**

Since the DS domains in the NC phase of 1T-TaS$_2$ must contribute to the out-of-plane metallicity, it follows that the same $\sqrt{13} \times \sqrt{13}$ reconstruction in the NC and C phases results in, respectively, metallic and insulating states. This is consistent with the results of recent density functional theory (DFT) calculations[23,24,32,33], indicating that the electronic band structure of commensurately distorted 1T-TaS$_2$ varies substantially depending on the interlayer stacking of DS, and that the insulating nature of the C phase is, in fact, a consequence of a special stacking order rather than Mott localisation. Moreover, the studies predicted significantly stronger energy-momentum dispersions along the ΓA direction of the Brillouin zone than in the ΓMK plane, along with favoured *c*-axis conduction, which has not been experimentally observed until now.

To assist our interpretation of the resistivity data we conducted DFT calculations for a different possible configurations of the commensurate superlattice. Given the strong electron-phonon and, possibly, electron-electron interactions in 1T-TaS$_2$, we enabled the relaxation of the lattice and included the on-site Coulomb energy *U*. In all of the previous theoretical works either one or both of these ingredients have been neglected. As in Lee et al.[24], we denote a configuration where for two successive layers the DS superlattices are vertically aligned with each other as the A stacking (Fig. 5a), and a configuration where the centre of a DS in one layer corresponds to an outermost Ta atom of a DS in a neighbouring layer (i.e. the DS avoid each other as much as possible) as the L stacking (Fig. 5b).

In the case of A stacking, the obtained band structure (Fig. 5c) has even lower dispersion in the ΓMK plane than suggested by the earlier studies and an effectively flat Fermi surface (Fig. 5e). According to the Boltzmann transport model, we can estimate the resultant resistivity anisotropy by considering the ratio of the Fermi-surface-averaged squares of the in-plane and out-of-plane Fermi velocity components. Astonishingly, the predicted value of $\rho_\perp/\rho_\parallel$ for the modelled state is approximately 0.023, while the analogous value for the completely undistorted lattice is as high as 40. The electronic structure of the L-stacked crystal is much more three-dimensional (Fig. 5d,f,g), but the density of states and the Fermi surface still have a pronounced quasi-one-dimensional character. Based on the orientation of the conductivity ellipsoid (Fig. 5g), the direction of maximal conductivity approximately follows the line connecting a DS to its nearest neighbour in an adjacent layer. The predicted ratios between the resistivity along the most conductive direction and the resistivities along the other two principal axes of the ellipsoid are around 0.077 and 0.2. The electronic dispersion along Γ-K-M-Γ path in Figure 5d is predominantly caused by the slanting of the "line-of-stars" (dashed line in Figure 5g) with respect to the *c*-axis. The effect of slanting, particular to the L stacking, should not be confused with a relatively minor effect of the in-plane DS hybridisation and its contribution to the in-plane conductivity.

The presented calculations indicate that for both types of stacking the commensurate DS superlattice develops a single-band quasi-one-dimensional dispersion in a wide window around the Fermi energy. This means that each DS effectively has a single electronic orbital, and that the nearest-neighbour tight-binding overlap between these orbitals is always significantly better for two DS's in successive layers than for a pair of DS's in the same layer. These overlaps form chains that follow the sequence of nearest-neighbour DS's in adjacent layers, with each DS having only one nearest-neighbour DS in any of the adjacent layers (plots of electron density throughout the A-stacked and L-stacked lattices are presented in the Supplementary Figure 12 and discussed in the Supplementary Note 5). Presence of these DS chains explains the coherent *c*-axis charge transport in 1T-TaS$_2$. In the NC phase the DS domains in adjacent layers partially overlap, and according to the published crystallographic data[14], in each overlap the DS's follow one of the three possible orientations of the L stacking (L, I and H types according to Lee at al.[24]). We therefore expect the orbital chains to exist in the NC phase in the regions where the domain overlap persists over a sizeable number of layers (see Fig. 1e). Given that these regions occupy only a fraction of the in-plane area, the intrinsic resistivity of the DS chains is likely even lower than the experimentally obtained value of $\rho_\perp$ for the NC phase. In the absence of structural dimerisation, and with sufficiently small amount of stacking disorder, the metallicity of the chains is expected to survive.

DFT calculations performed without the inclusion of the Coulomb energy $U$ produced electronic structures with slightly more pronounced three-dimensional characters (presented in the Supplementary Note 6 and the Supplementary Figure 13) yet the overall picture remains qualitatively the same as in the finite $U$ case.

**Metal-insulator transition in 1T-TaS$_2$ as a quasi-one-dimensional instability**

The proposed formation of *c*-axis oriented conducting chains in the NC phase has interesting implications in the context of the metal-insulator transition in 1T-TaS$_2$. Specifically, the NC to C phase transition can be regarded as a quasi-one-dimensional instability, driven by the energy gain from the modulation of the interlayer hybridisation of the DS orbitals. The mechanism resembles the Peierls instability – a metal-insulator transition occurring as a result of a dimerisation of a one-dimensional chain. However, such a comparison is very superficial given the continuous nature of the order parameter in genuine Peierls systems. In 1T-TaS$_2$ the change of the CDW modulation occurs in a discontinuous manner, through a discrete change of DS stacking, which explains the first order type of the phase transition. The proposal to view the insulating state of the C phase as a consequence of the particular stacking order has recently been made by Ritschel et al.[23] and Lee at al.[24] The idea has been discussed with a meticulous comparison of cohesive energies and bands structure calculations for various stacking orders, and it decisively challenged the long standing Mott localisation scenario. However, there has so far been no direct experimental observation of the out-of-plane-metallic state in the NC phase of 1T-TaS$_2$, serving as a precursor in the stacking-order localisation picture. This gap is conveniently filled by our resistivity data, revealing the *c*-axis metallicity in the NC phase. Simultaneously, our DFT calculations show that the interlayer DS hybridization is a major consequence of structural relaxations, fully unconstrained with respect to the positions of atoms within the supercell, and the lattice constants in all directions. The interlayer DS hybridization also shows independently of the imposed Bravais lattices types, choice of which leads to spontaneous formations of the different types of DS stacking.

## Discussion

To summarise, by employing the state-of-the-art FIB-based sample preparation process, we conducted reliable and accurate measurements of the in-plane and out-of-plane electrical resistivities of 1T-TaS$_2$ in multiple CDW phases. The experiments revealed unusually anisotropic charge transport character in presence of the NC CDW order: the out-of-plane conduction is metallic, yet for the in-plane direction the material becomes less conductive upon cooling. We point out the substantial enhancement of the in-plane resistivity, compared to the typical values observed in other metallic TMDs, and conclude that the aperiodic and composite structure of the phase is likely at the root of this effect. At the same time, the out-of-plane conduction is favoured by the formation of the quasi-one-dimensional *c*-axis oriented chains of electronic orbitals, propagating through the overlapping regions of DS domains and serving as a source of previously unobserved coherent c-axis metallicity – a description supported by our DFT calculations. Consequently, the layered material has an uncharacteristically low resistivity anisotropy, exhibiting nearly 5 times lower out-of-plane resistivity than the in-plane one under certain conditions. Moreover, we interpret the NC to C phase transition as a quasi-one-dimensional instability, and support the idea of the C phase being a band insulator, rather than a Mott state, as has been thought previously. Our findings are also of value to the efforts aimed at functionalising the compound. Exploiting the interlayer conduction nearly doubles the resistivity change upon the low temperature phase switching, compared to the difference seen in the in-plane charge transport, leading to a more robust performance of a potential device. Furthermore, miniaturisation and production scalability requirements additionally favour utilisation of the out-of-plane charge transport. 1T-TaS$_2$ serves as a neat example of how interlayer interactions in a van der Waals material lead to unexpected properties, warranting a more systematic study of resistivity anisotropy in other compounds.

## Methods

**Synthesis of 1T-TaS$_2$ single crystals.** The single crystals of 1T-TaS$_2$ were prepared from the elements using the chemical vapour transport technique via a reversible chemical reaction with iodine as a transport agent, between 950 °C (hot zone) and 900 °C (cold zone). The 1T-polytipic phase is obtained by the addition of SnS$_2$ (less than 0.5% in weight) and by rapid cooling from the growth temperature[47].

**Focused ion beam microfabrication.**

Samples of 1T-TaS$_2$ used for this study were extracted from monocrystalline flakes of approximately 100 µm thickness and the lateral size of the order of 1 mm. In order to produce the sample shown in Fig. 2a, a Helios G4 Xe Plasma FIB microscope, manufactured by FEI, was first used for isolating a rectangular lamella of 115 µm length, 60 µm width and 4 µm thickness from the single crystal by milling away the surrounding material. For this process the ion column voltage and the beam current were set respectively to 30 kV and 60 nA. The 60 µm long edges of the lamella were aligned with the $c$-axis of the crystalline lattice. After the lamella was extracted, its surface was polished via grazing angle milling at a significantly lower beam current of 1 nA. This procedure ensured the parallelism of the two largest faces of the sample. The lamella was subsequently positioned on a sapphire substrate with the largest face parallel to the surface. The exposed surfaces of the lamella and the substrate were etched by argon plasma for 5 minutes and sputter-coated by a 5 nm titanium layer followed by a 150 nm gold layer at a 20° angle to the surface normal. We draw attention to the fact that no adhesive was used for attaching lamella to the substrate in order to reduce the effects of differential thermal contraction and compressibility during the subsequent experiments. The remaining FIB procedures were conducted in Helios G2 Ga FIB microscope (also produced by FEI). Pt contact points were grown by gas assisted deposition. Clearly visible in Fig. 2a of the main text of the article, they acted both as electrical connections and anchoring points, securing the lamella on the substrate. The deposited material is amorphous and has a higher resistivity than a pure metal, therefore in order to make the final electrodes more conductive the specimen was sputter-coated by the second 100 nm layer of gold. The ion beam was then used to expose the surface of 1T-TaS$_2$ by milling away the metal film from the surface of the desired current paths of the lamella. Using a beam current of the order of 10 nA the current channel was defined by milling a set of trenches. The current channel had three extended straight sections between 3 µm and 5 µm wide, two parallel and one perpendicular to the $c$-axis of the crystal. Each of these sections had two branching points at which the voltages were probed. In order to improve the mechanical robustness of the structure, the concave edges were rounded. As a final step, all exposed faces of the sample along the current channel were finely polished with the ion beam of 1 nA current in order to reduce the surface roughness to the level of nanometers and remove most of the contaminated or damaged surface layer. The final dimensions of the sample, used for converting resistance to resistivity, were determined from the scanning electron microscope images.

**Resistivity measurements.** Resistivity was measured via the four-point technique with direct or alternating excitation currents in the 20–40 µA range. Temperature sweeps were conducted with the maximum rate of 1 K/min for the ambient pressure and of 0.5 K/min for the high-pressure measurements in order to avoid generating substantial temperature gradients.

A hydrostatic high-pressure environment for resistivity measurements was created using a piston cylinder cell produced by C&T Factory. 1:1 pentane-isopentane mixture was used as a pressure-transmitting medium (remains liquid at room temperature for up to approximately 6 GPa[48]). Pressure was determined from the changes in resistance and superconducting transition temperature of a sample of Pb located next to the 1T-$TaS_2$ sample.

**Powder X-ray diffraction measurements.** The high-pressure powder X-ray diffraction data were collected at the BM01 Swiss-Norwegian beamline of the European Synchrotron Radiation Facility (ESRF) in Grenoble. The beam had a wavelength of 0.7458 Å and a 100 × 100 nm size. PILATUS@SNBL detector[49] was used for signal acquisition and the sample-detector distance was 146 mm. High pressures were generated with a diamond anvil cell using Daphne oil 7373 as the pressure-transmitting medium. Stainless steel gasket was pre-indented to 50 µm thickness and had a hole of 250 µm diameter. Pressure was determined from the shift in the fluorescence wavelength of ruby[50]. The sample was rotated by 4° to improve statistics. Integrations were done with only minimal masking using BUBBLE software[49]. CrysAlis PRO and JANA2006[51] software packages were used for data processing and structural refinement, respectively.

**Density functional theory calculations.** The density functional theory calculations were performed using the Quantum ESPRESSO package[52], with ultrasoft pseudopotentials from Pslibrary[53]. The kinetic energy cut-off for wave functions was 70 Ry, whereas the kinetic energy cutoff for charge density and potential was 650 Ry. We used PBE exchange energy functional[54] and Marzari-Vanderbilt smearing[55] of the Fermi surface of 0.001 Ry. The Brillouin-zone sampling used in self-consistent calculations was 6×6×12 k-points (with no shift). The Fermi surface were calculated with a denser mesh of 12×12×24 k-points. The on-site Coulomb interaction on Ta ions was taken into account within the DFT+$U$ approach proposed by Cococcioni and de Gironcoli[56]. The Hubbard interaction $U$ on Ta atoms was set to 2.41 eV. This choice of $U$ is somewhat arbitrary and was guided by the values for $U$ in 1T-TaS$_2$ suggested by other authors[31,57] ($U$ = 2.27 eV and $U$ = 2.5 eV), as well as the value ($U$ = 2.94 eV) that we obtain for the undeformed 1T-TaS$_2$ structure using the appropriate module within Quantum ESPRESSO [58,59]. The proper determination of $U$ requires the self-consistent procedure where the calculation of the optimal structure is accompanied by the continuous re-evaluation of $U$[60,61]. For DS-based superstructures of 1T-TaS$_2$ this is bound to produce a list of $U$'s with somewhat different value for each inequivalent Ta positions within the crystal. The convergence thresholds used for atomic structure relaxation were 0.0001 (a. u.) and 0.001 (a. u.) for the total energy and atomic forces, respectively. The DFT calculations without DFT+$U$ corrections were made in the same manner.

## Data availability

The data that support the findings of this study are available from the authors (E.M. and K.S.) upon reasonable request.

## Acknowledgements

We would like to express gratitude to Dr. Tobias Ritschel (TU Dresden), Prof. Philipp Aebi (University of Fribourg), Dr. Nicholas Ubrig (University of Geneva) and Prof. Dr. Andrew Mackenzie (MPI CPfS Dresden) for valuable discussions and feedback, as well as Dr. Gaetan Giriat (EPFL) for the instrumentation related support and Dr. Maja Bachmann (MPI CPfS Dresden) for her assistance with FIB microfabrication. We also appreciate the allocation of the beamtime at the Swiss–Norwegian Beam Lines (SNBL) by the SNX council and thank Dr. Vladimir Dmitriev and the BM01 staff for their support and assistance during the X-ray diffraction experiments. This study has been funded the Swiss National Science Foundation through its SINERGIA network MPBH and grant No. 200021_175836. The work by I.B. was supported in part by Croatian Science Foundation Project No. IP-2018-01-7828. The work by E.T. was supported in part by Croatian Science Foundation Project No. IP-2016-06-7258. A. Akrap acknowledges funding from the Swiss National Science Foundation through Project No. PP00P2_170544. C.P. and P.J.W.M. acknowledge the support from the Max Planck Society.

## Author contributions

E.M. and K.S. performed a part of the FIB microfabrication, prepared and conducted resistivity measurements at ambient and high pressures, analysed the resistivity data and wrote the manuscript. A.P. performed the early iterations of FIB microfabrication and together with L.C prepared and conducted the first resistivity measurements at ambient and high pressures. I.B and E.T. performed the DFT calculations and provided a theoretical support. C.P. performed a part of the FIB microfabrication and together with P.J.W.M. provided a FIB-related expertise. The X-ray diffraction experiments were conducted by A. Arakcheeva and L.C. and the gathered data were analysed by A. Arakcheeva. A. Akrap advised on various aspects of the work and helped with the manuscript preparation. H.B. synthesised the single crystals of 1T-TaS$_2$. The idea behind the study was conceived by L.F., who also headed the laboratory hosting the study. The project was supervised by L.F. and K.S.

## Competing interests

The authors declare no competing interests.

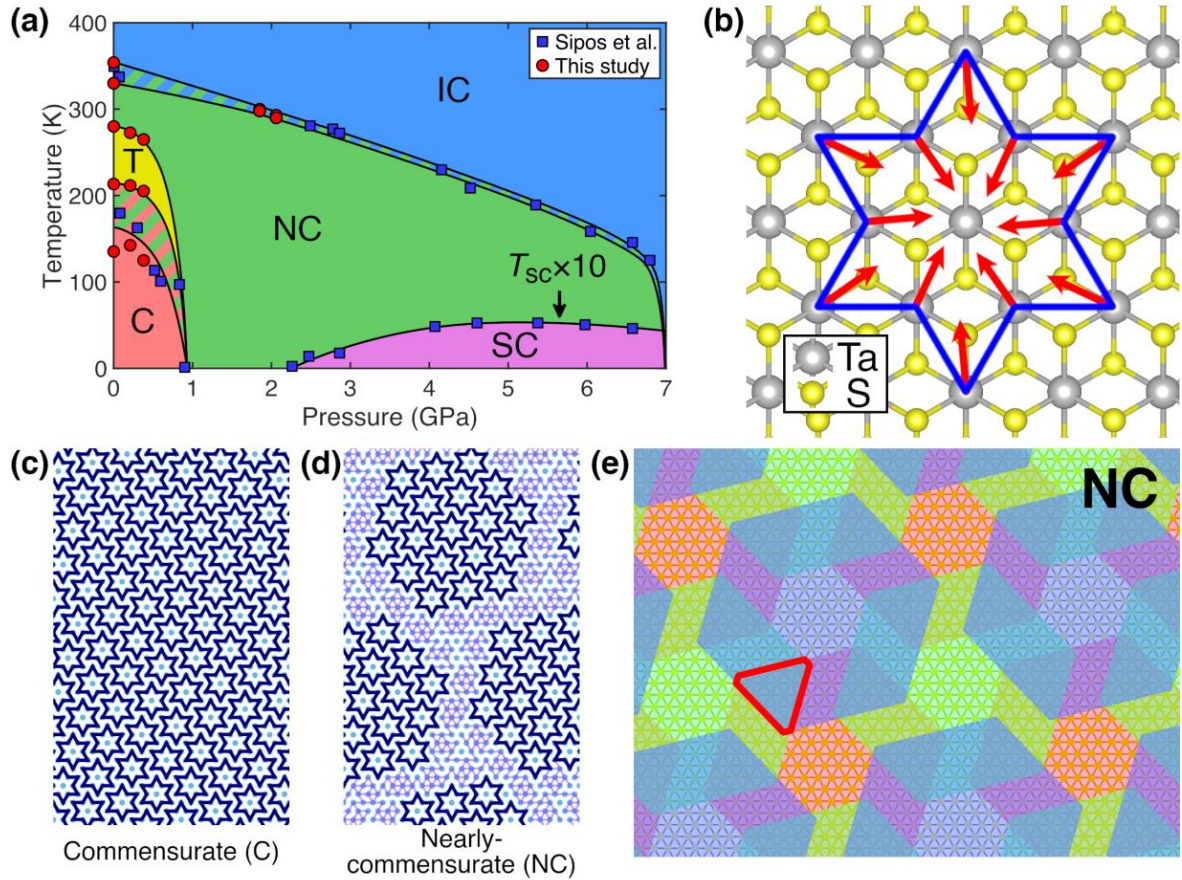

**Figure 1 | Lattice distortion in different phases of 1T-TaS$_2$. a,** Pressure-temperature phase diagram of 1T-TaS$_2$ based on the resistivity data of Sipos et al.[12] and this study. The labels stand for commensurate (C), nearly-commensurate (NC), triclinic (T) and incommensurate (IC) charge density wave phases, as well as the superconducting (SC) phase. The black lines represent the approximate phase boundaries. The striped areas indicate the regions where hysteresis occurs upon heating and cooling. The triclinic phase is observed only during a warm-up. Superconducting transition temperature is multiplied by 10 for clarity. **b,** Visualisation of the in-plane displacement of Ta atoms (red arrows) leading to the formation of a 13-atom David-star-shaped cluster (blue outline). **c,d,** Schematic illustration of the in-plane lattice distortions in the commensurate (**c**) and nearly-commensurate (**d**) charge density wave phases. The David-star clusters are marked with the dark blue outlines. Light blue dots stand for the Ta atoms. The purple mesh represents locally aperiodic reductions of Ta-Ta distances. **e,** Schematic visualization of the *ab*-plane projection of the interlayer stacking of the reconstructed domains in the NC phase. Blue, red and greed hexagons represent the domains in three consecutive layers. The red outline marks a region where the domains overlap throughout the whole *c*-axis period.

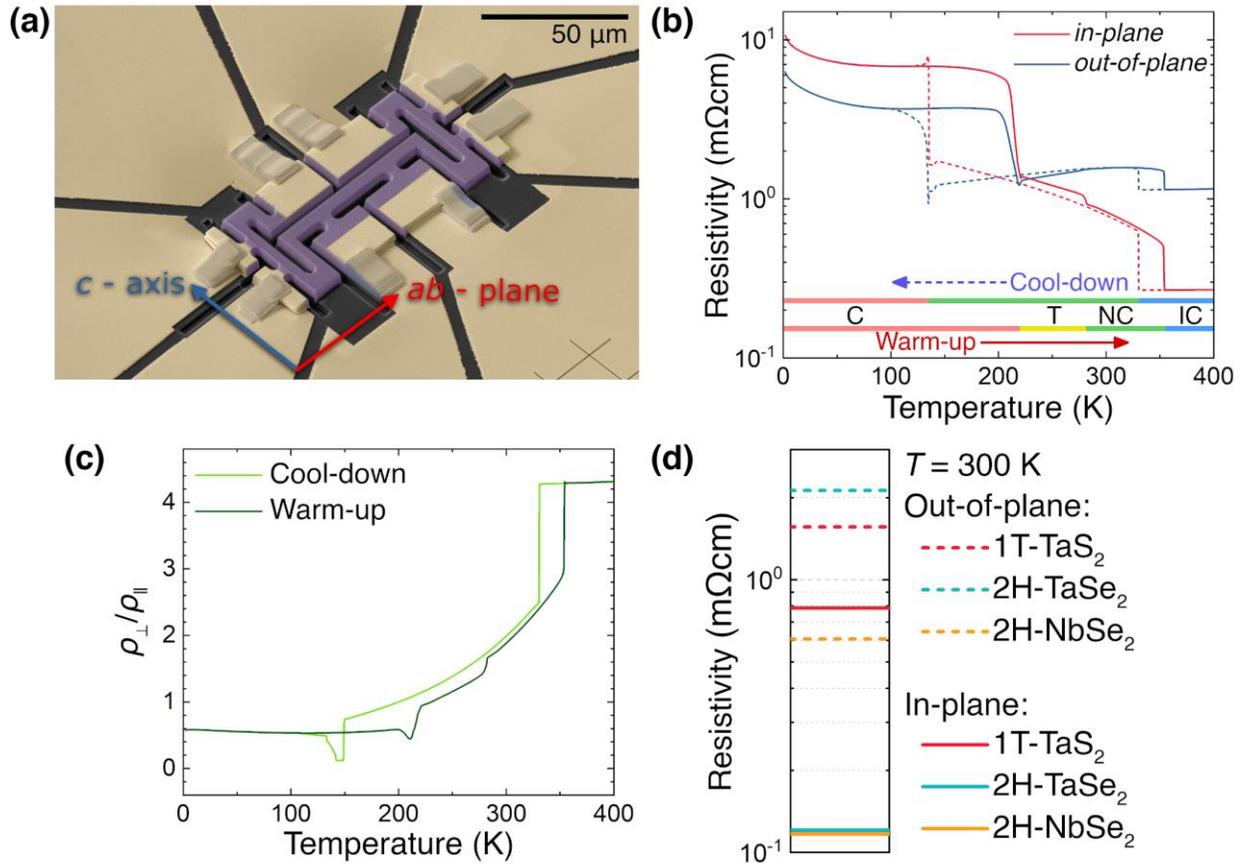

**Figure 2 | Focused-ion-beam-fabricated sample and electrical resistivities of 1T-TaS$_2$, 2H-TaSe$_2$ and 2H-NbSe$_2$. a,** Scanning electron microscope image of the bulk monocrystalline sample of 1T-TaS$_2$ prepared using focused ion beam. False colouring is used for illustrative purposes. The arrows indicate the orientation of the crystal (purple). A layer of gold (beige) formed the conductive paths between the crystal and external leads. The rectangular ramps, created using in-situ platinum deposition, provide mechanical stability and improve the continuity of the gold layer. The scale bar in the top-right corresponds to a 50 μm distance. **b,** Temperature dependence of electrical resistivities of 1T-TaS$_2$ measured on the microfabricated sample. The bar below the curves indicates temperature intervals corresponding to the different phases encountered upon heating (solid line) and cooling (dashed line). Sharp features accompanying the jump in resistivity near 135 K appear due to temporary internal stresses experienced by the sample as a result of a rapid volume change during the phase transition. **c,** Ratio of interlayer to intralayer resistivities ($\rho_\perp/\rho_\parallel$) of 1T-TaS$_2$ as function of temperature. **d,** Comparison between the room temperature in-plane and out-of-plane resistivity values of 1T-TaS$_2$ and two other metallic transition metal dichalcogenides: 2H-TaSe$_2$ and 2H-NbSe$_2$. All samples were microstructured with focused ion beam.

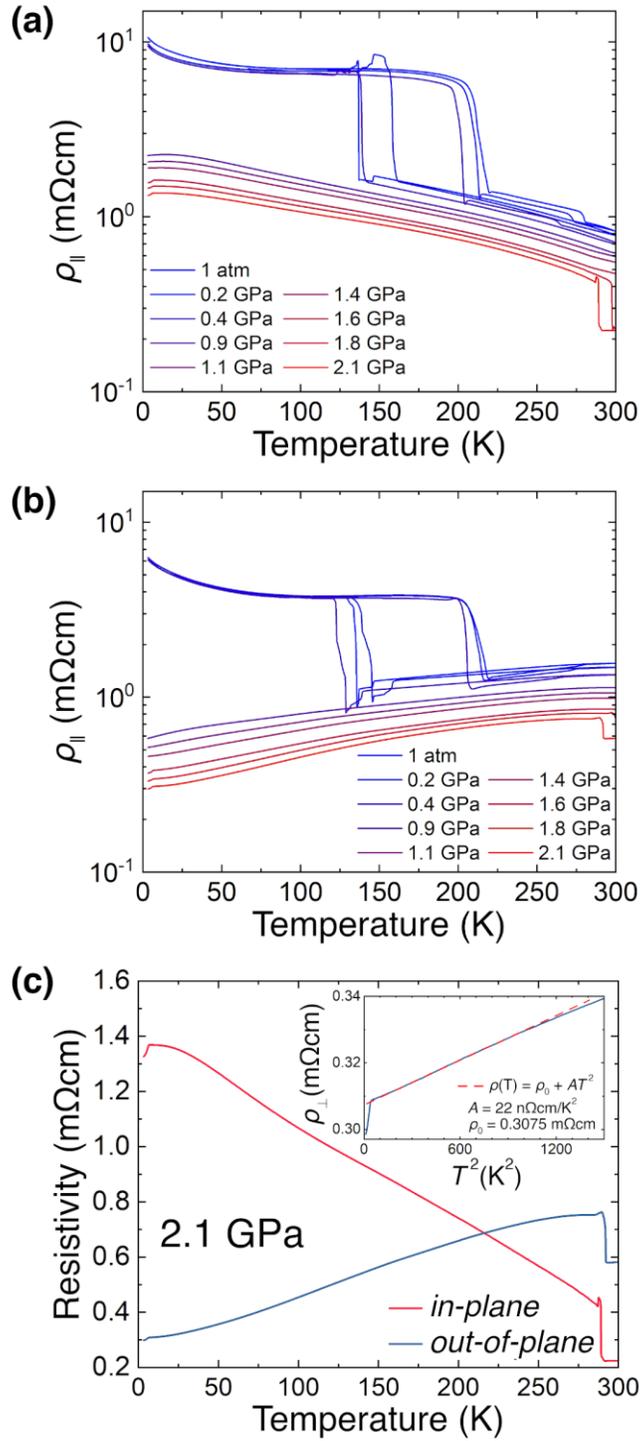

**Figure 3 | High-pressure resistivity of 1T-TaS$_2$. a,b,** In-plane (**a**) and out-of-plane (**b**) electrical resistivities of 1T-TaS$_2$ ($\rho_\parallel$ and $\rho_\perp$, respectively) as functions of temperature for different applied hydrostatic pressures. **c,** Temperature dependences of resistivities along the two directions at 2.1 GPa pressure. The inset shows a quadratic power-law fit to the out-of-plane resistivity between 7 K and 30 K.

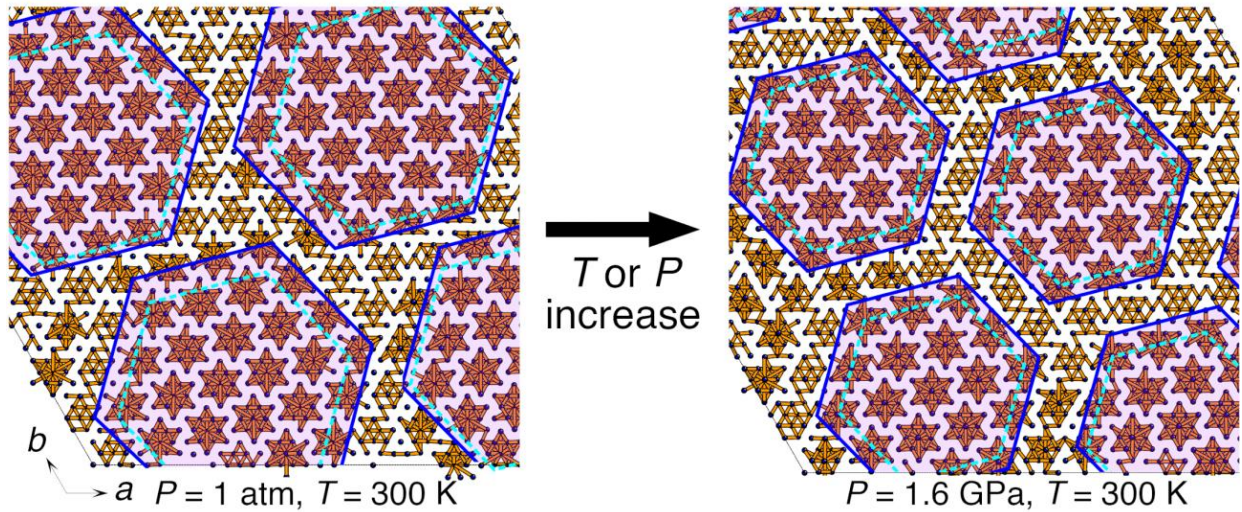

**Figure 4 | Domain lattice rotation in 1T-TaS$_2$.** Room temperature in-plane lattice structures in the nearly-commensurate phase of 1T-TaS$_2$ at ambient pressure (left) and at 1.6 GPa (right), visualised based on the powder X-ray diffraction data. Ta atoms are represented by dots, which are connected if the interatomic separation is below an arbitrarily chosen threshold value. The hexagonal outlines mark the commensurately reconstructed domains with the incomplete David-star-shaped clusters included (solid blue) or excluded (dashed cyan). Temperature ($T$) and pressure ($P$) increases both individually lead to the same transformation – rotation from corner-sharing towards side-sharing arrangement, shrinkage of domains and reduction of the domain lattice period.

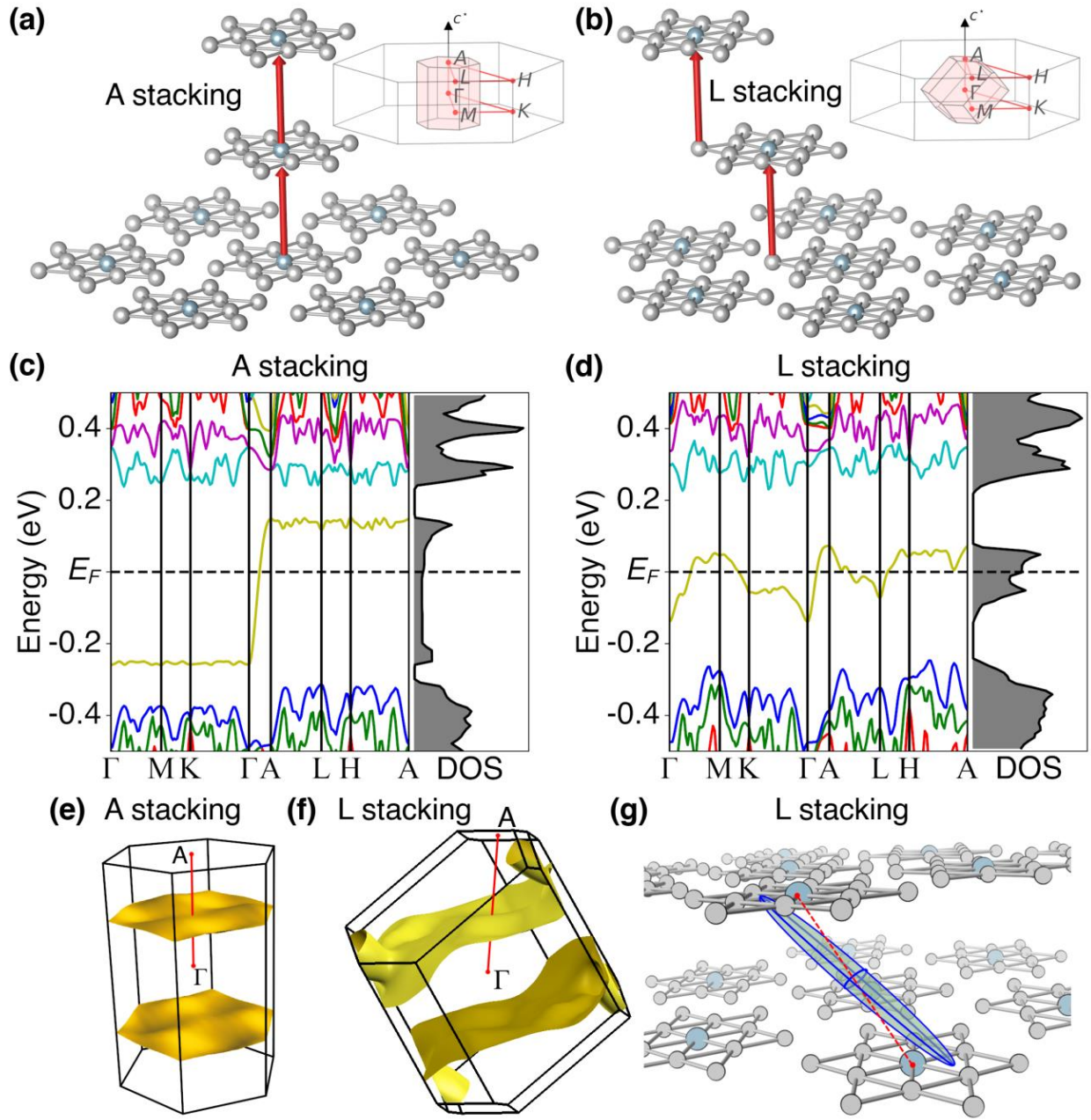

**Figure 5 | Numerical modelling of the electronic structure of 1T-TaS$_2$. a,b,** Arrangements of the David-star-shaped clusters (only Ta atoms are shown) in successive layers of commensurately distorted 1T-TaS$_2$, considered for our density functional theory (DFT) calculations. After Ref. 24, we denote these stacking configurations as A (**a**) and L (**b**). The red arrows join the Ta atoms sharing the same lateral position. The top right sections of both panels display the first Brillouin zones for the corresponding lattices in relation to the first Brillouin zone for the undistorted 1T-TaS$_2$. **c,d,** The respective electronic band structures and densities of states (DOS) predicted by the DFT calculations for relaxed lattices and with the on-site

electron-electron interaction energy $U$ = 2.41 eV. The directions traced in the reciprocal space are marked in the panels **a** and **b**. Each colour stands for a different band. **e,f,** The respective Fermi surfaces. The extended open sheets in both cases indicate a quasi-one-dimensional character of the electronic structures. **g**, Conductivity ellipsoid calculated from the electron energy dispersion near the Fermi surface (i.e assuming isotropic relaxation time), in relation to the crystal structure for the L stacking. The longest axis of the ellipsoid indicates the direction of the highest conductivity. The ellipsoid is slightly slanted (approximately 8°) with respect to the line connecting the centers of two nearest-neighbour clusters in two adjacent layers. The ratios of the highest conductivity to the conductivities along the two shorter principal axes of the ellipsoid are around 13 and 5.

**Supplementary Information for**

**Preferential out-of-plane conduction and quasi-one-dimensional electronic states in layered 1T-TaS$_2$**

E. Martino[*,1,2], A. Pisoni[1], L. Ćirić[1], A. Arakcheeva[1], H. Berger[1], A. Akrap[2], C. Putzke[3,4], P. J.W. Moll[3,4], I. Batistić[5], E. Tutiš[6], L. Forró[1], K. Semeniuk[*,1]

[1]*École Polytechnique Fédérale de Lausanne (EPFL), Institute of Physics, CH-1015 Lausanne, Switzerland*
[2]*University of Fribourg, Department of Physics, CH-1700 Fribourg, Switzerland*
[3]*École Polytechnique Fédérale de Lausanne (EPFL), Institute of Materials Science and Engineering, CH-1015 Lausanne, Switzerland.*
[4]*Max Planck Institute for Chemical Physics of Solids, 01187 Dresden, Germany.*
[5]*Department of Physics, Faculty of Science, University of Zagreb, HR-10000 Zagreb, Croatia*
[6]*Institute of Physics, HR-10000 Zagreb, Croatia.*

*Corresponding authors: edoardo.martino@epfl.ch (E.M.), konstantin.semeniuk@epfl.ch (K.S.).

**Supplementary Note 1. Microstructuring layered crystals using focused ion beam**

Due to a weak mechanical interlayer coupling, characteristic for many layered compounds, the final microstructure is extremely vulnerable to delamination due to shear stresses. Throughout our study mechanical failures of this kind resulted in several FIB-structured samples getting irreversibly damaged, particularly during the high pressure experiments (Supplementary Figure 1). We therefore emphasise the importance of minimizing the shear stresses potentially experienced by the microstructure. The details of the fabrication process, such as choosing the particular dimensions, not having any adhesion between the probed region of the sample and the substrate and rounding of the convex edges, were crucial ingredients for the successful measurements.

It also needs to be kept in mind that small size of a FIB-made sample enhances the Joule heating experienced by it. For the dimensions used in this study, excitation current of the order of milliamperes is capable of completely destroying narrow parts of the sample due to excessive heating. We therefore limited the excitation current used in our measurements to the maximum value of about 50 μA.

Exposing a single crystal to FIB unavoidably subjects surface to some damage, penetrating to a depth of about 10–20 nm. This phenomenon generally does not manifest in resistivity measurements[1], but the notion should still be considered, since it could have been, in principle, possible that during our measurements the intrinsic interlayer charge transport properties of 1T-$TaS_2$ had been obscured by a contribution from a thin shell of damaged compound. We addressed this question by measuring the out-of-plane resistivity on two parts of the crystal of different cross-section areas (within the same microstructure). Given different surface-to-volume ratios of the two channels, we found that there is no significant difference between their respective out-of-plane resistivity values, implying that the surface damage had a negligible effect.

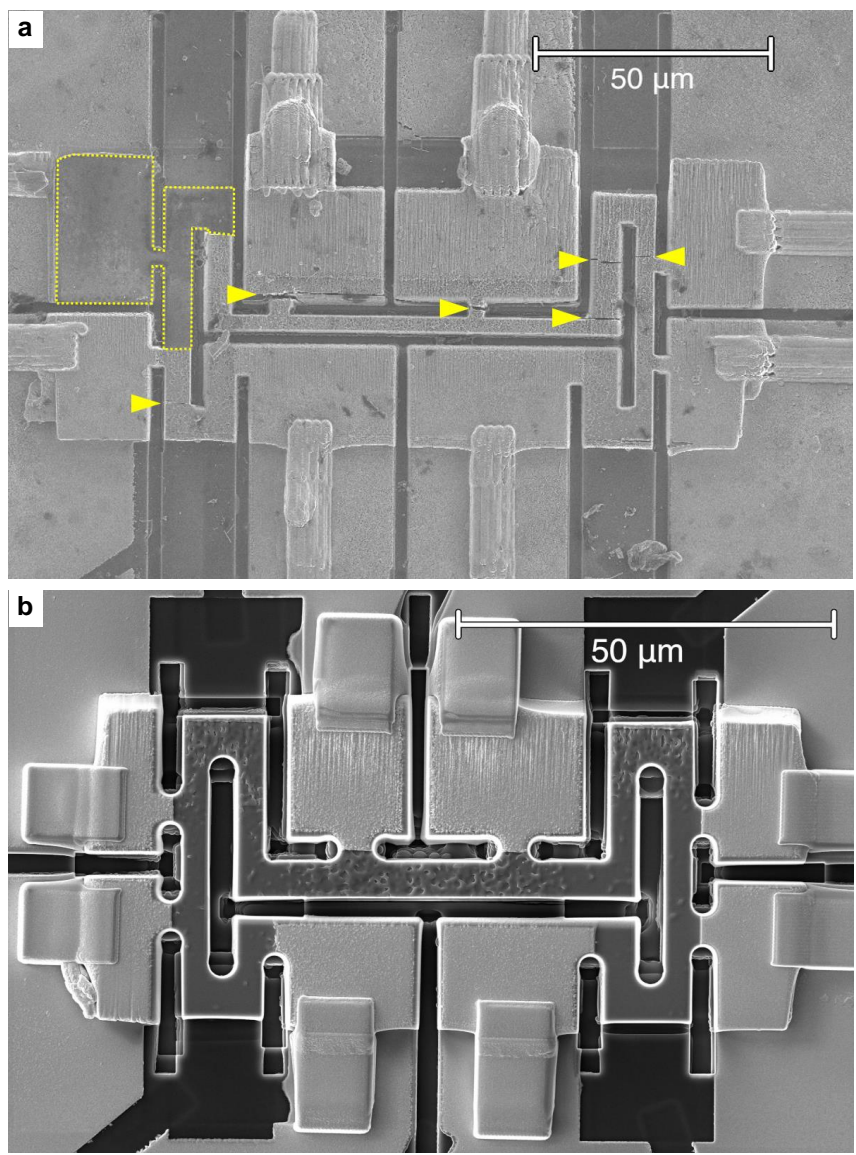

**Supplementary Figure 1 | Optimisation of sample geometry for high-pressure experiments.**
**a**, Scanning electron microscope image of a microstructured 1T-TaS$_2$ sample, damaged during a high-pressure experiment. Multiple cracks (marked by yellow arrowheads), propagating along the planes of layers, can be seen near convex edges of the sample, where mechanical stresses are enhanced. A missing portion of the sample is outlined with yellow dashed line **b**, Image of the microstructured 1T-TaS$_2$ sample of the latest iteration. In order to reduce the likelihood of mechanical failures, the minimum feature size was increased and all the convex edges were rounded. The sample successfully survived pressurisation to 2 GPa (image taken before applying pressure). In both images the *c*-axis of the crystal is oriented vertically.

**Supplementary Note 2. Validation of resistivity measurements by finite elements simulations**

To prove the reliability of our experimental results and assess the applicability of the chosen sample geometry for resistivity anisotropy measurements, we used the finite element analysis package COMSOL Multiphysics$^{TM}$ in order to calculate the current density and voltage fields in a model replicating our microstructured sample (Supplementary Figure 2). We imposed a fixed in-plane resistivity $\rho_\parallel$ = 0.78 mΩcm (room temperature value for 1T-TaS$_2$) and investigated how the resistivity anisotropy calculated from the potential differences across the voltage probes compares to the actual anisotropy, dictated by the variable out-of-plane resistivity $\rho_\perp$. As can be seen in Supplementary Figure 3, for anisotropy values between approximately 0.1 and 100 we expect measurements to provide reliable values of $\rho_\parallel$ and $\rho_\perp$.

Supplementary Figure 4 shows the computed current density field inside the sample for the anisotropy values of 0.1, 1 and 100. It can be seen that for the highest and lowest values of $\rho_\perp/\rho_\parallel$ the current jetting effects result in a slight divergence of the current flow lines along some of the probed sections of the crystal. In the isotropic case, which, according to our data, is the closest to the real situation, the current flows homogeneously between every pair of voltage probes.

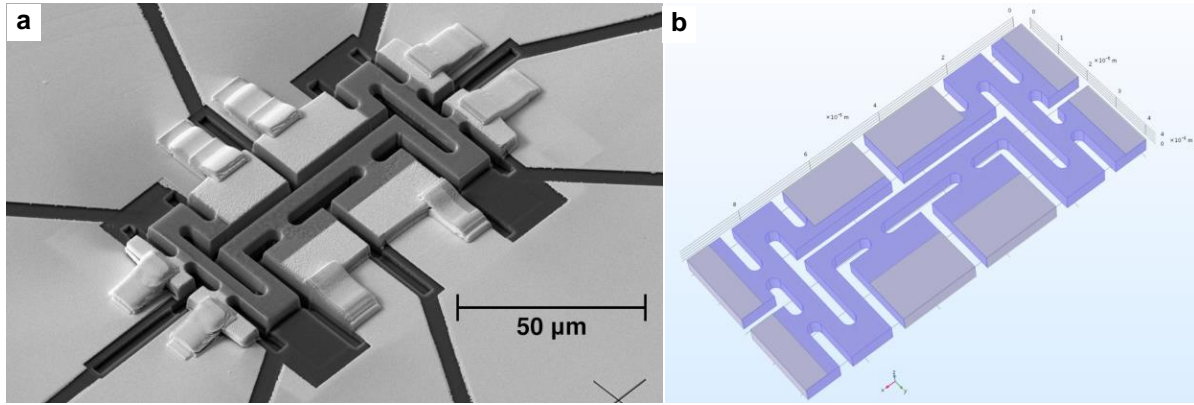

**Supplementary Figure 2 | Focused-ion-beam-structured sample. a**, A scanning electron microscope image of the focused-ion-beam-structured sample of 1T-TaS$_2$ used for our study (see Fig. 2a of the main paper for the explanation of various features). **b**, A virtual model of the sample created for conducting finite element simulations of the current flow. Blue — 1T-TaS$_2$, grey — gold layer. The dimensions of the actual sample and the model match.

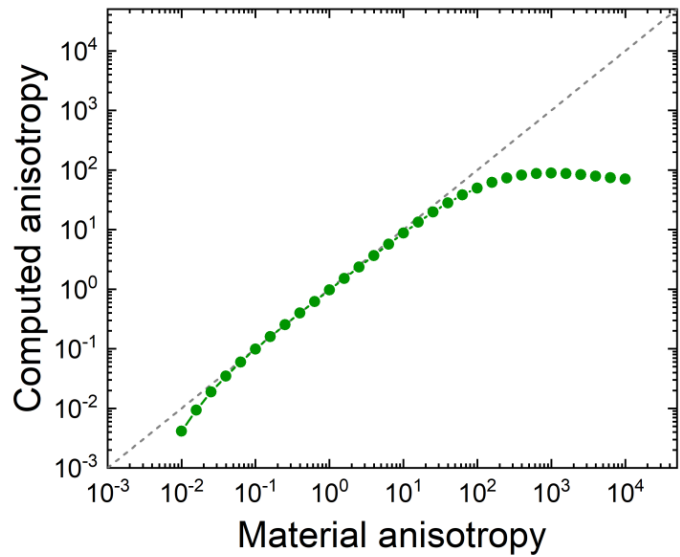

**Supplementary Figure 3 | Applicability of the chosen sample geometry for resistivity anisotropy measurements.** Finite element calculations were used to compute voltages at the probing points of the microstructured sample (see Supplementary Fig. 1), which were used to calculate the apparent resistivity anisotropy (Computed anisotropy). The computed anisotropy is plotted (green dots) against the imposed intrinsic anisotropy (Material anisotropy). The dashed line indicates the perfect agreement between the values of the two anisotropies. The chosen sample geometry can be used to reliably measure the out-of-plane and in-plane resistivities if the ratio of the between them is roughly between 0.1 and 100.

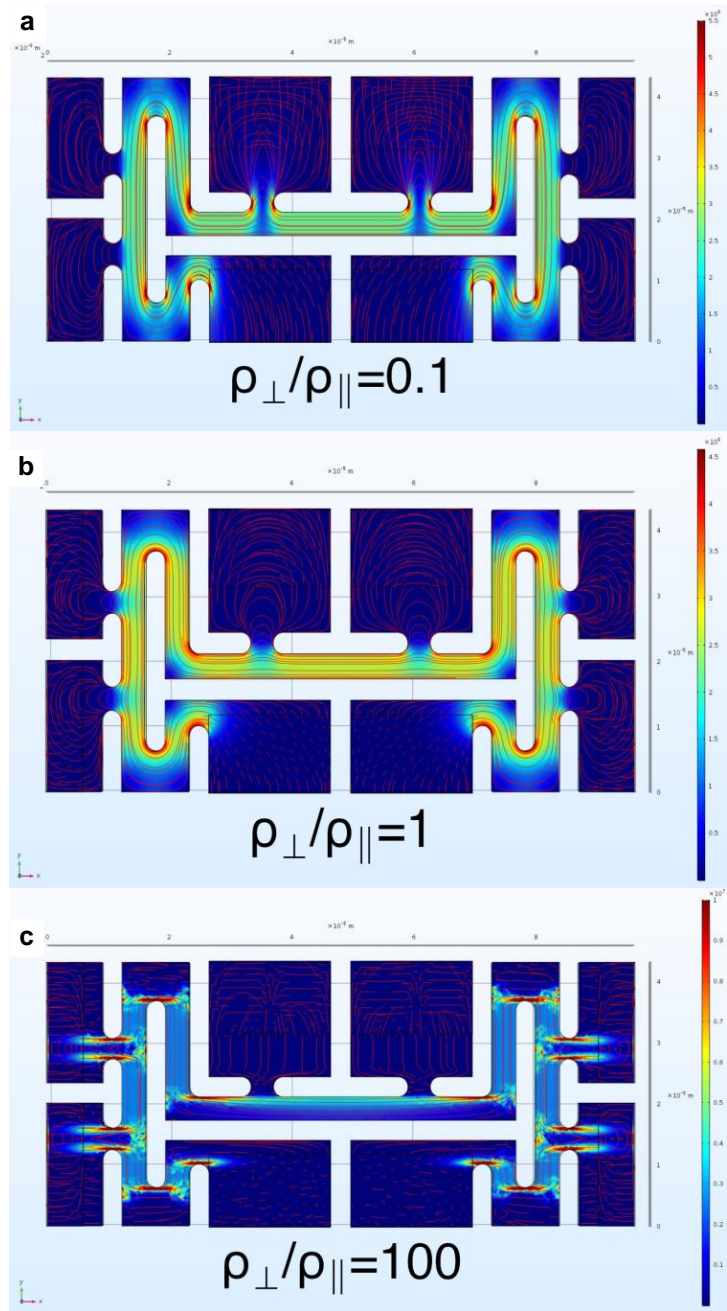

**Supplementary Figure 4 | Simulated current density field in a microstructured sample.** The plots show the magnitude of the current density (colour map) and the current flow trajectory (red lines) predicted by the finite element calculations for the focused-ion-beam-structured sample (see Supplementary Fig. 1). The calculations were done for three different values of resistivity anisotropy ($\rho_\perp/\rho_\parallel$): 0.1 (**a**), 1 (**b**) and 100 (**c**). The *ab*-plane and *c*-axis are aligned, respectively, with the horizontal and vertical directions. In the $\rho_\perp/\rho_\parallel = 1$ case the current flows homogeneously along all the probed sections of the sample. For $\rho_\perp/\rho_\parallel = 0.1$ small current jetting can be observed in the *c*-axis oriented parts of the sample. For $\rho_\perp/\rho_\parallel = 100$ some inhomogeneity in the current density can be seen inside the in-plane channel.

**Supplementary Note 3. Explaining inconsistencies with the previously published 1T-TaS$_2$ resistivity anisotropy values**

The comparable $\rho_\parallel$ and $\rho_\perp$ of 1T-TaS$_2$, presented in the main text of the article, are in an obvious contradiction with the $\rho_\perp/\rho_\parallel$ values of approximately 500 and 2000 obtained, respectively, in the experiments of Hambourger & Di Salvo[2] and of Svetin et al.[3] In this section we provide the likely explanations of these discrepancies.

In the case of Svetin et al. resistivity measurements have been performed on a very thin flake of 1T-TaS$_2$ (the dimensions of 110 µm x 10.5 µm x 90 nm are provided in the paper), which has been probed via 8 electrodes, with 4 located on each side of the flake (Supplementary Figure 5a). The out-of-plane resistance has been measured via the two-point method, with subsequent determination and subtraction of the contact resistances. When converting resistance to resistivity, the authors assumed that the current was distributed across the whole area of the flake, which would have been correct had the resistivity anisotropy of 1T-TaS$_2$ been much more than 1. According to our results, this is not the case, therefore the much higher $\rho_\perp$ can be explained by the massive overestimation of the current carrying cross-section area.

Another issue is the fact that the plots of the $\rho_\parallel$ and $\rho_\perp$ against temperature provided in the paper by Svetin et al. have almost the same shape. Given the small thickness of the sample, if the contacts used for the interlayer resistivity measurements have been misaligned by more than several tens of nanometers, the resultant current path must have had a substantial in-plane component. Voltage drop due to this contribution could have been large enough to dominate over the genuine out-of-plane signal, resulting in effectively the same quantity being measured for the two nominally different current flow directions.

The experiment of Hambourger & Di Salvo utilised a slightly unconventional measurement configuration, illustrated in Supplementary Figure 5b. Using the notation introduced in the figure, and defining $V_{BB'}$ and $V_{CC'}$ as the potential differences between contacts B and B' as well as C and C', respectively, the authors claimed that resistivity anisotropy should have been inversely proportional to $(\ln V_{BB'} - \ln V_{CC'})^2$. For low enough anisotropy most of the current is expected to flow only in a small portion of the sample, in close vicinity of the contacts A and A', making the voltages $V_{BB'}$ and $V_{CC'}$ extremely small and therefore difficult to measure. In order to verify this statement we used the finite element analysis package COMSOL Multiphysics$^{TM}$ for calculating the distributions of current density and voltage throughout the sample during a measurement.

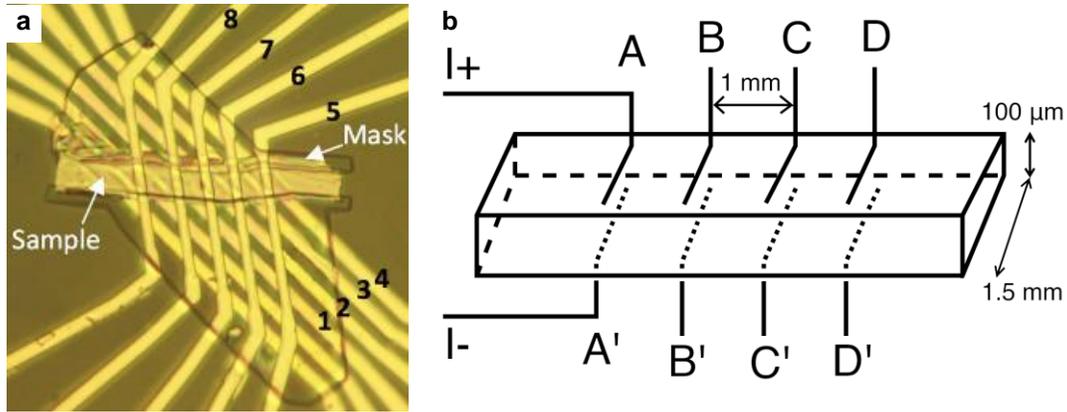

**Supplementary Figure 5 | Measurement geometries of the earlier studies of the resistivity anisotropy of 1T-TaS$_2$. a**, Photograph of the 1T-TaS$_2$ sample and the measurement electrodes from the paper of Svetin et al.[3] The *c*-axis of the crystal is perpendicular to the plane of the image. The electrodes 1–4 and 5–8 run, respectively, below and above the sample. **b**, Schematic representation of the measurement geometry used in the experiment of Hambourger & Di Salvo[2] (redrawn according to the diagram in the paper). The smallest dimension of the sample (100 µm) indicates the *c*-axis extent of the 1T-TaS$_2$ crystal.

Supplementary Figure 6 shows the model used for the calculations. Following the information provided in the paper, the sample was represented by a rectangular block of 5 mm length, 1.5 mm width and 100 µm thickness (*c*-axis extent). The spacing between the electrodes was 1 mm and the electrodes themselves were chosen to be 50 µm wide. The excitation current was set to 5 mA, the same value as quoted in the paper.

The virtual sample was assigned the in-plane resistivity value equal to that of 1T-TaS$_2$ at room temperature (0.78 mΩcm). Imposing resistivity anisotropy of 10 results in the current density and voltage distributions shown in Supplementary Figure 7. As expected, effectively all the current flowed directly between the electrodes A and A', making $V_{BB'}$ and $V_{CC'}$ less than the resolution limit of 1 nV. Such a weak signal could have been easily overshadowed by various contributions due to systematic errors. A set of calculations for various anisotropy values (Supplementary Figure 8) shows that for $\rho_\perp/\rho_\parallel < 100$, voltage $V_{BB'}$ becomes more than two orders of magnitude smaller than the potential difference between the current electrodes A and A' (voltage $V_{CC'}$ is substantially smaller than $V_{BB'}$), meaning that barely any current flows close to the electrodes B and B'. The given measurement geometry is, therefore, only applicable for highly anisotropic materials.

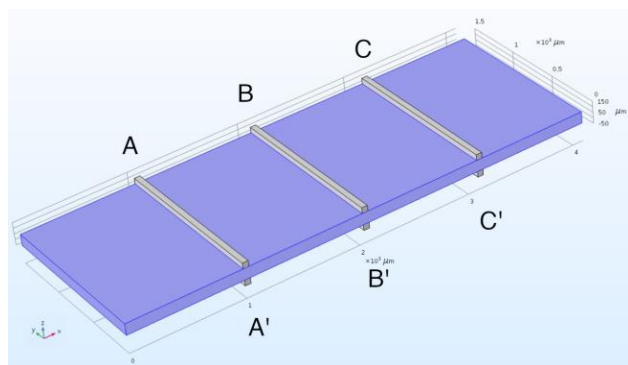

**Supplementary Figure 6 | Model of a 1T-TaS$_2$ sample for simulating an earlier resistivity anisotropy measurement.** A virtual model of the 1T-TaS$_2$ sample (blue) in the experiment of Hambourger & Di Salvo[2] (see Supplementary Fig. 4b) used for finite element calculations. Grey bars play the role of the current (A, A') and voltage (B, B', C, C') electrodes. The dimensions (specified in the text) are consistent with the information provided in the paper.

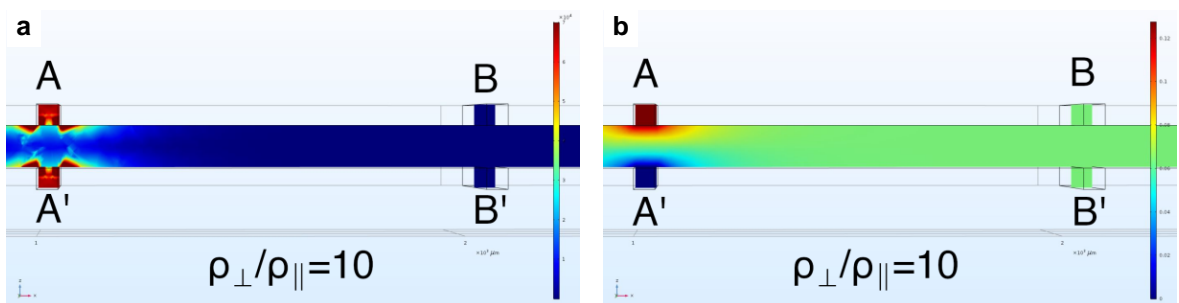

**Supplementary Figure 7 | Simulated current density and voltage for an earlier 1T-TaS$_2$ resistivity anisotropy experiment.** Colour plots of the calculated current density (**a**) and voltage (**b**) inside a sample for the measurement geometry used in the study of resistivity anisotropy ($\rho_\perp/\rho_\parallel$) of 1T-TaS$_2$ by Hambourger & Di Salvo[2]. The view corresponds to a plane parallel to the longest and the shortest edges of the sample and passing through the centre of it (see Supplementary Fig. 5 for reference). The *c*-axis of the crystal is in the vertical direction. Even for the imposed anisotropy of 10 (an order of magnitude higher than in reality) the current is concentrated in the vicinity of the current electrodes A and A', leaving the voltage sensing electrodes B and B' effectively at the same potential.

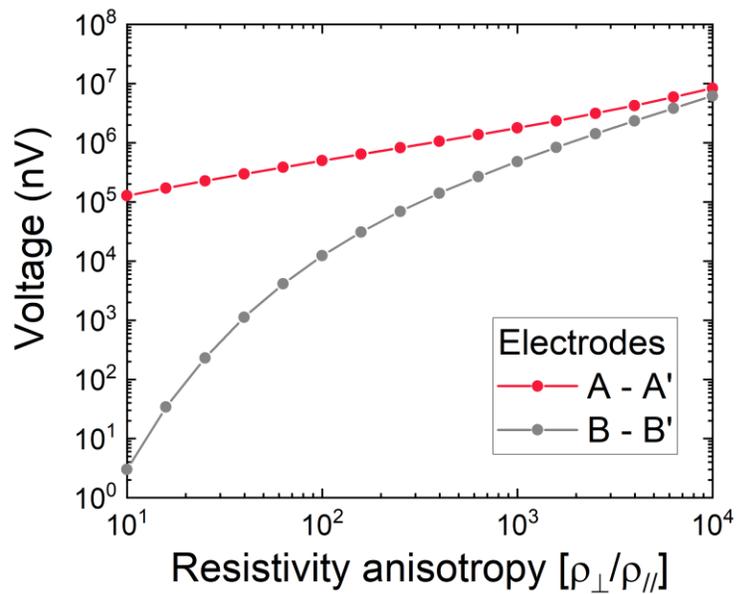

**Supplementary Figure 8 | Predicted voltage across electrodes as function of resistivity anisotropy.** Potential differences across pairs of electrodes A and A' as well as B and B' (Hambourger & Di Salvo experiment[2], see Supplementary Fig. 5b and Supplementary Fig. 6), predicted by finite element calculations, plotted as functions of resistivity anisotropy.

**Supplementary Note 4. Powder X-ray diffraction measurements**

The complexity of the lattice distortion in the NC phase of 1T-TaS$_2$ is an important ingredient in determining the associated charge transport properties. Detailed information regarding the real-space structural parameters of the NC phase is scarce in the literature, therefore we conducted a synchrotron X-ray diffraction study of powdered 1T-TaS$_2$. The measurements were conducted at room temperature and at pressures of up to 2.1 GPa.

The minimally processed X-ray diffraction data can be seen in Supplementary Figure 9. Supplementary Figure 10 shows how the average intralayer and interlayer Ta-Ta distances, as well as the volume of the crystal vary between 1 atm and 2.1 GPa. For this range of pressures the changes are approximately linear with about 5 times more rapid compression along the *c*-axis than along the layers. To further quantify the fragility of the C phase, we note that the pressure of 0.9 GPa, required for its full suppression, corresponds to merely 1.3% reduction in the *c* lattice constant.

A thorough analysis showed that the in-plane superstructure modulation of the NC phase can be described by the wavevector $\boldsymbol{q} = \alpha \boldsymbol{a}^* + \beta \boldsymbol{b}^*$ (where $\boldsymbol{a}^*$ and $\boldsymbol{b}^*$ are the reciprocal lattice vectors and $\alpha$ and $\beta$ are scalar coefficients). The obtained values of $\alpha$ and $\beta$ are consistent with the published data from a high pressure study of 1T-TaS$_2$ single crystals[4]. Monotonic variation of the coefficients upon pressure increase corresponds to the rotation of the $\boldsymbol{q}$-vector towards $\boldsymbol{a}^*$, while its magnitude remains at a constant value of 0.284(1) Å$^{-1}$ (Supplementary Figure 11). This corresponds to the rotation of the DS domain lattice upon the application of pressure from an approximately corner-sharing configuration at 1 atm towards a honeycomb-like arrangement of hexagons at 1.6 GPa (see Figure 4 of the main text). The edges of the DS domains maintain their orientation (13.9° with respect to the *a*-axis of the unreconstructed lattice) while the periodicity and size of the domains decrease. This real-space depiction of the lattice distortion in the NC phase is significantly more intricate than a commonly used representation by a Kagome lattice of hexagonal DS domains. The observed rotation closely resembles the structural change happening as a function of temperature, discovered in early scanning tunnelling microscopy experiments[5,6]. The domains cannot shrink indefinitely, and a further pressure increase results in a discontinuous transition to the IC phase during which $\beta$ abruptly goes to zero while $\alpha$ increases, keeping $|\boldsymbol{q}|$ constant.

The refinement allowed us to extract additional details regarding the structural evolution of the NC phase of 1T-TaS$_2$ with pressure. The following changes occur upon increasing pressure from 1 atm to 1.6 GPa at 293 K:

- The average Ta-Ta distance for the first/second nearest neighbour within a DS decreases from 3.37 Å/5.84 Å to 3.34 Å/5.80 Å.
- The extent of a DS domain along its longest dimension, including/excluding incomplete clusters, goes from 5/4 to 4/3 clusters.
- The percentage of Ta atoms involved in the clusters, including/excluding incomplete ones, goes down from 52%/81% to 30%/50%.
- The period of the domain lattice is reduced from 75 Å to 65 Å.
- The average distance between the clusters along the *c*-axis changes from 5.9 Å to 5.8 Å.

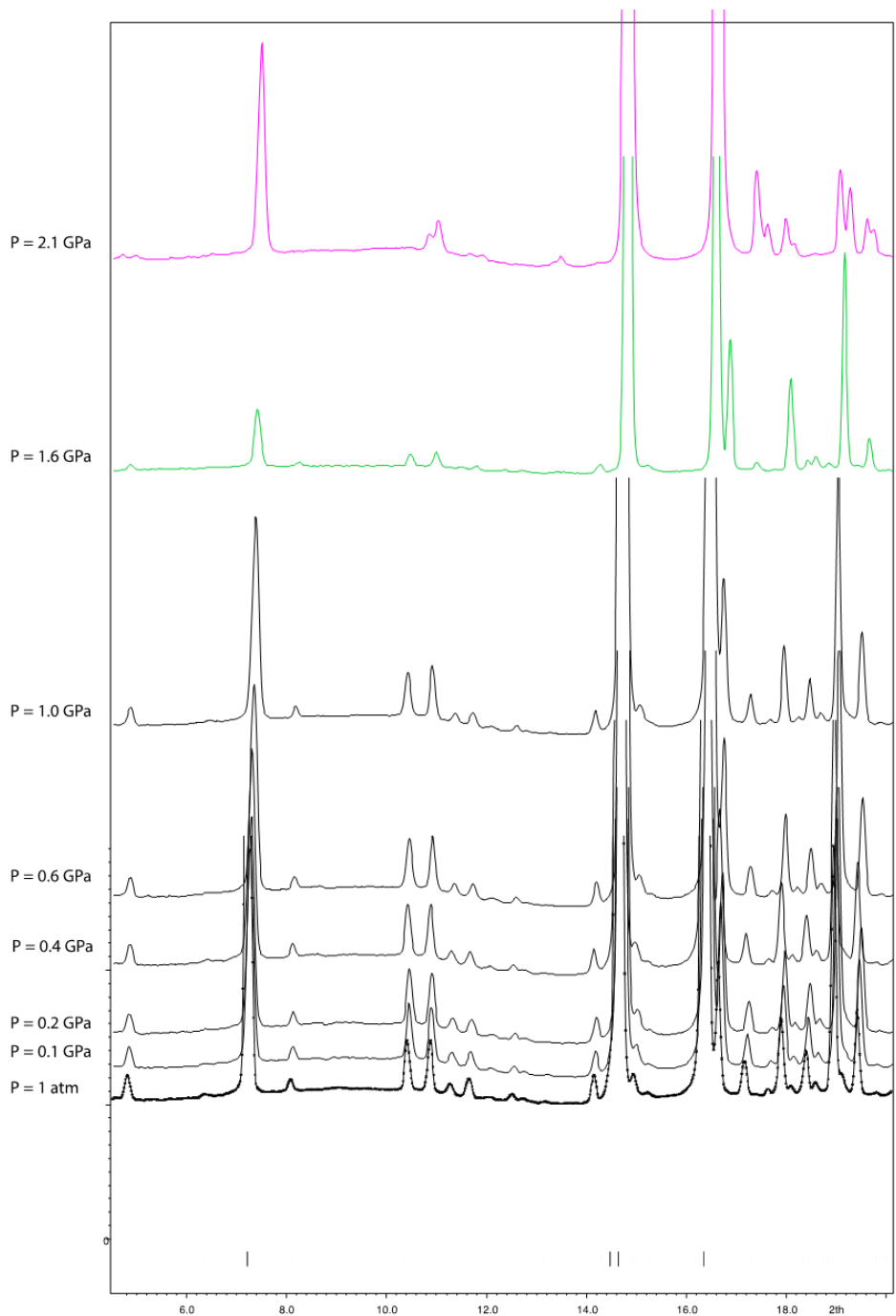

**Supplementary Figure 9 | High pressure powder X-ray diffraction data for 1T-TaS$_2$.** Plots of intensity against the angle between the incident and reflected beams for different pressures. Measurements were done at room temperature. The ticks at the bottom of the plot indicate the Bragg reflections corresponding to the hexagonal unit cell parameters $a \approx 3.36$ Å and $c \approx 5.8$ Å. The other reflections are satellites of the superstructure modulation, with the wavevector $\boldsymbol{q} = \alpha \boldsymbol{a}^* + \beta \boldsymbol{b}^*$ (see the text for the definitions of symbols).

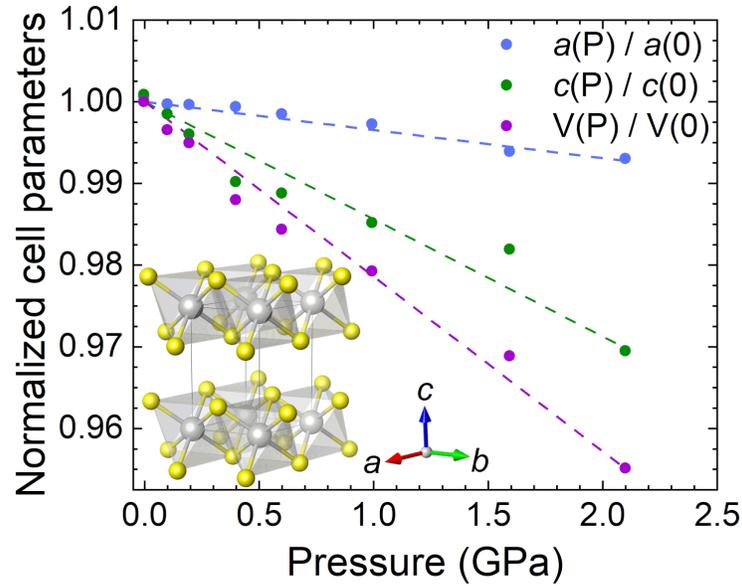

**Supplementary Figure 10 | Compressibility of 1T-TaS$_2$.** Pressure dependence of the mean Ta-Ta distances within (*a*) and between (*c*) layers as well as the overall volume change (*V*). For the given pressure range the changes can be approximated as linear, which is illustrated with the fitted dashed lines. The crystal structure is visualised in the bottom left (grey – Ta, yellow – S).

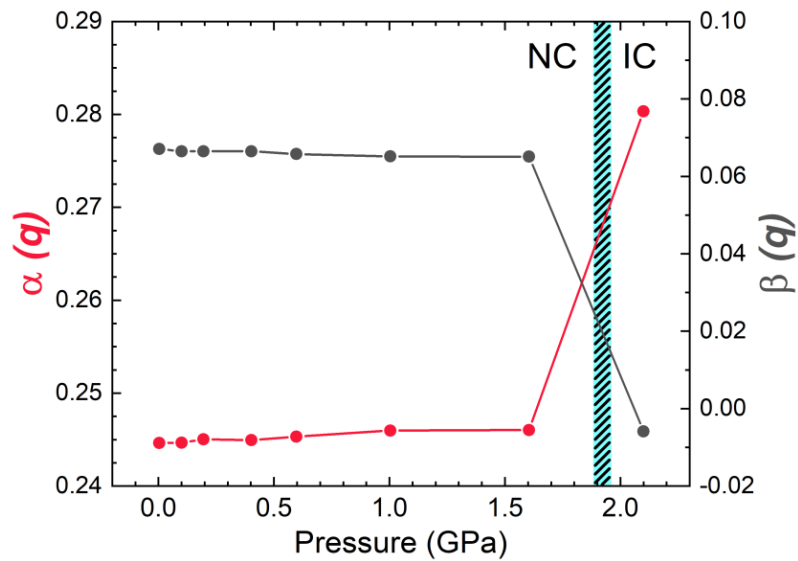

**Supplementary Figure 11 | Change of the lattice distortion of 1T-TaS$_2$ with pressure.** Pressure dependence of the coefficients $\alpha$ and $\beta$ of the in-plane lattice distortion modulation wavevector $\boldsymbol{q} = \alpha \boldsymbol{a}^* + \beta \boldsymbol{b}^*$ (see the text for the definitions of symbols). The vertical stripe at 1.9 GPa marks the phase boundary between the NC and IC phases.

**Supplementary Note 5. Real-space visualisation of the David-Star chains via electron density plots**

The formation of DS chains through interlayer hybridization appears to be the main mechanism beyond distinct intra-layer and inter-layer transport properties of 1T-TaS$_2$ in the NC phase. At the level of geometry, the appearance of chains is the consequence of the structure of DS layers. This structure imposes that each DS in a given layer has only one nearest-neighbour DS in each of the adjacent layers, irrespective of the stacking. The precise form of the chains depends on the type of stacking and its periodicity. In the cases of A and L stackings, addressed in our DFT calculations, the DS chains follow straight lines. The NC phase has a 3-layer periodicity, so one can expect the DS chains to follow a helicoidal path with the stacking following a repeating sequence of L, I and H types for each consecutive pair of layers. At the level of electron dynamics, the chains appear as the consequence of strong hybridization within a DS—resulting in a formation of a single effective DS orbital—and a better hybridization between DS's located in different adjacent layers than between DS's in the same layer. These two features are directly visible in the electron density, calculated from the occupied states of the conduction band, shown in Supplementary Figure 12 for the A and L stackings. First, within the layers the charge density in the regions between DS's is insignificant in comparison to the charge density inside a DS (Supplementary Figures 12a,b). Second, the charge density in the space between a pair of nearby DS's in two neighbouring layers is visibly larger that the charge density between pairs of DS's in the same layer (Supplementary Figures 12c,d).

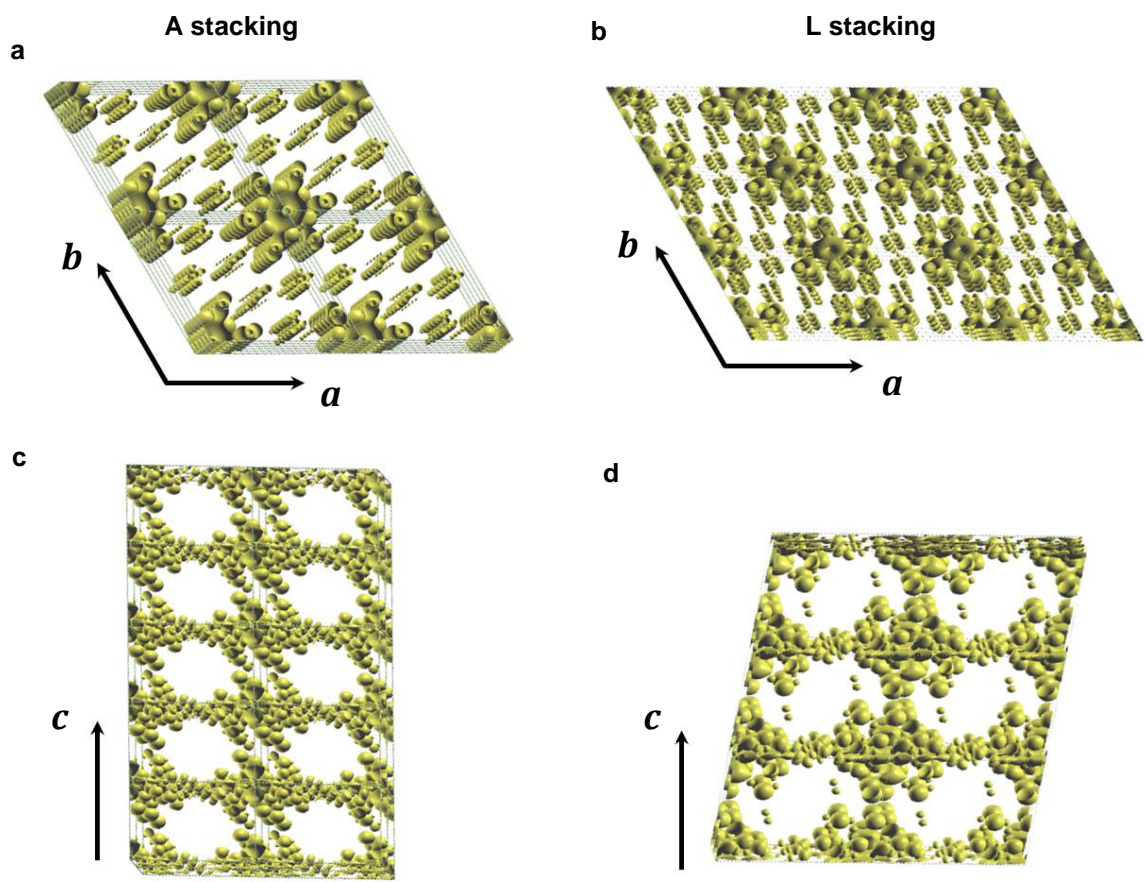

**Supplementary Figure 12 | David-star chains illustrated through charge density in fully relaxed structures.** The spatial distribution of the electron density, illustrated though iso-density surfaces[7], and calculated from the occupied states in the conduction band for relaxed structures of the A-stacked (**a, c**) and L-stacked (**b, d**) commensurately distorted 1T-TaS$_2$. The distributions are viewed along the directions almost parallel (**a, b**) and almost perpendicular (**c, d**) to DS chains. The plots contain 2×2×5 DS supercells for the A stacking, and 3x3x3 DS supercells for the L stacking. The arrows indicate the principal crystallographic directions.

**Supplementary Note 6. The DFT results without DFT+*U* corrections**

Here we graphically present the DFT results obtained in the absence of the DFT+*U* correction. Supplementary Figure 13 shows the electronic dispersion and the Fermi surfaces for the A-stacking (panels a, c) and the L-stacking (panels b, d), obtained by fully relaxing the crystal structure without the DFT+*U* corrections. These figures can be compared to the corresponding figures obtained for *U* = 2.41 eV, shown in Figs. 5c-f in the main text. The conduction bands get significantly narrower without the *U* correction, whereas the quasi-one-dimensional anisotropy in conductivity is approximately 50% smaller. The shape of the Fermi surface in the absence of the DFT+*U* corrections remains very much the same for the A-stacking. For the L-stacking the Fermi surface remains open, but with much more pronounced three-dimensional features.

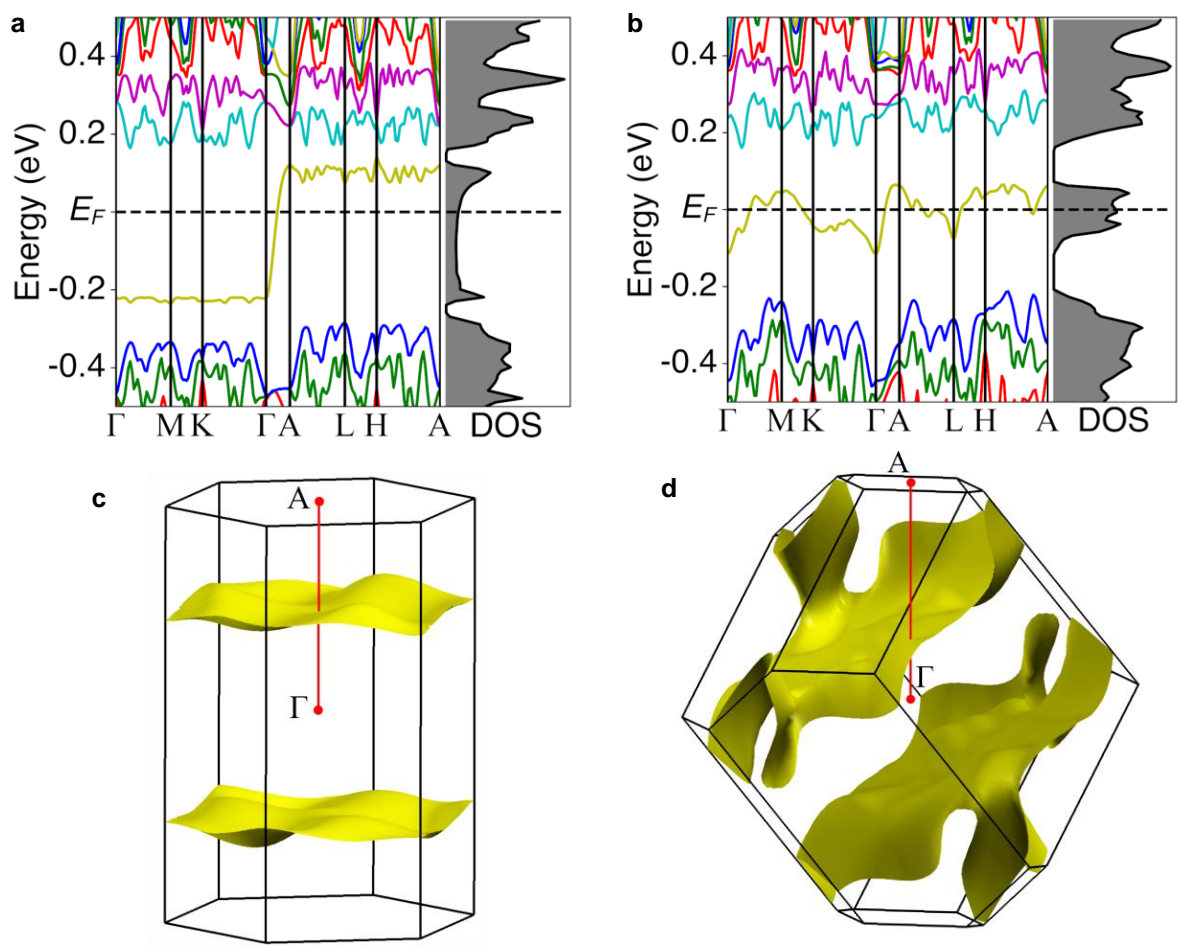

**Supplementary Figure 13 |** The electronic dispersion and the Fermi surfaces for A-stacking (**a** and **c**) and L-stacking (**b** and **d**) obtained by fully relaxing the crystal structure without DFT+*U* corrections.

**Supplementary References**